\documentstyle[preprint,aps]{revtex}

\begin{document}
\title{Direct Observation of Superheating and Supercooling of Vortex Matter
using Neutron Diffraction}
\author{X. S. Ling,$^{1}$ S. R. Park,$^{1}$ B. A. McClain,$^{1}$ S. M.
Choi,$^{2,3}$ D. C. Dender,$^{2,3}$
 and J. W. Lynn$^{2}$}
\address{$^1$Department of Physics, Brown University, Providence, Rhode
Island 02912\\
$^{2}$NIST Center for Neutron Research, Gaithersburg, Maryland 20899\\
$^{3}$University of Maryland, College Park, Maryland 20742\\
}
\maketitle



\ \\
\abstract{We report the first observation of a striking history dependence of the
structure function
of the vortex matter in the peak effect regime in a Nb single crystal by
using small angle neutron scattering combined with {\it in situ} magnetic
susceptibility
measurements. Metastable phases of vortex matter, supercooled vortex liquid
and superheated vortex solid,
have been identified.  We interpret our results as direct structural
evidence for
a first-order solid-liquid transition at the peak effect.\ \\
\ \\
PACS numbers: 74.60.Ge, 61.12.Ex}

\pacs{74.60.Ge, 61.12.Ex}

 A current subject of wide interest concerns the existence of a vortex
solid-liquid transition
in type-II superconductors \cite{blat}.  In addition to providing a model
system for the study of
melting and freezing,
the vortex matter system offers unprecedented opportunities in studying the
effects of quenched
disorder on phase transitions. Recently,
the magnetization jump in high-$T_c$ superconductors
YBa$_2$Cu$_3$O$_{7-\delta}$ (YBCO),
a widely accepted thermodynamic signature of a vortex solid-liquid
transition \cite{zel},
and the peak effect, a well-known vortex-lattice softening anomaly
in many low-$T_c$ \cite{kes,bha} and high-$T_c$ \cite{lin,kwo,ish,shi}
superconductors in which the critical
current exhibits a peak rather than decreasing monotonically with
increasing temperature, are found to occur at the same temperature
\cite{ish,shi}.
 However, there is no direct structural evidence indicating whether the
underlying phase transition is solid-to-solid, solid-to-liquid,
or even liquid-to-liquid in origin.
Since quenched disorder is known to have important consequences
for phase transitions \cite{lar,ffh}, whether a solid-liquid
transition can occur when random pinning is effective
 has broad implications in condensed matter physics.

Small angle neutron scattering (SANS) \cite{cri,cub,kei,lyn,gam} is presently
the most powerful technique for probing the vortex structure in bulk
superconductors.
For high-$T_c$ superconductors, the magnetic penetration depth is typically
quite long, yielding
a very weak vortex signal in SANS experiments, which has so far prevented a
direct structural
study of the peak effect in high-$T_c$ systems.
Recent effort has therefore focused on low-$T_c$ systems where one can
detect the peak effect
and measure the vortex structure using SANS
in the same temperature and field regime \cite{lyn,gam}, and
for these purposes Nb is experimentally the most favorable system because
of the short penetration depth and the availability of large, ultrahigh
quality single crystals.
However, it has been controversial as to whether there is a
vortex solid-liquid transition in Nb and the two previous SANS studies of
this issue
were not conclusive
\cite{lyn,gam}.
 In the first study \cite{lyn}, a characteristic temperature at which the
azimuthal width
  of the Bragg peaks starts to increase was interpreted as the
 vortex-lattice melting transition.
 In the second study \cite{gam}, a nearly isotropic ring of scattering
 was observed in the peak-effect regime, but was interpreted as a
disordered solid
 rather than a liquid.  In both studies, only field-cooled vortex states
were measured.
 We report here striking hysteresis
 in neutron diffraction patterns of the vortex state in the peak-effect
regime of Nb.
The metastability of supercooled and superheated vortex states
is directly demonstrated by applying a weak perturbation and observing the
evolution of the SANS patterns.  These data provide direct evidence
for a first-order vortex solid-liquid transition.

Our experiments were performed using
 the 30-m SANS instrument NG-7 at the NIST
Center for Neutron Research.
 The increased flux on the SANS instrument due to a new liquid-hydrogen
cold source
 turns out to be important, but the key improvement provided by the present
experiment is that
the SANS and the ac magnetic
response of the vortex
array are measured simultaneously {\em in situ}.
Thus one can correlate the ordering in the vortex state
directly with the macroscopic vortex dynamics.
The coil for the magnetic
ac susceptibility measurements can also be used to apply a small ac
field to dynamically perturb the vortex system, allowing an {\em in situ}
determination of the metastability of the vortex states.

The sample is a Nb single crystal of 99.998$\%$ purity, a cylinder with
a slightly uneven diameter from
1.316 cm at one end to 1.169 cm at
the other,
and 2.48 cm in length.
The incident
neutron beam has a mean wavelength $\lambda $=6.0 {\AA } and a bandwidth $%
\Delta \lambda $/$\lambda $=0.11. The neutron beam traverses through the
central region
of the sample,
defined by a cadmium mask (0.7 cm in diameter), along the cylindrical axis
which coincides with the three-fold symmetric $<$111$>$ crystallographic
direction.
The dc magnetic field is applied by a horizontal
superconducting magnet along the same direction. The absolute accuracy of
the measured sample
temperature is $\pm $ 0.20 K with a temperature stability better than $\pm
$ 0.025 K.

The peak-effect regime of the Nb sample is
determined {\em in situ} by measuring the characteristic dip in the
temperature dependence of the
real-part of the ac magnetic susceptibility $\chi'(T)$, as shown in
Fig.1(a) for $H$=3.75 kOe. The pronounced diamagnetic dip in $\chi'(T)$ of
the ac susceptibility
corresponds to a strong peak effect in the critical current (or nonlinear
conductance)
in the sample \cite{lin,shi}.  The onset, the peak, and the end of the peak
effect
are denoted by
 $T_o(H)$, $T_p(H)$, and $T_{c2}(H)$, respectively.
 (In YBCO, the end of the peak effect is still far below
$T_{c2}(H)$\cite{lin,kwo,ish,shi}.)
Fig.1(b) shows the window of our experiment. The upper critical field
$H_{c2}$(4.20K) = 4.23 kOe
of this sample is higher than
that of Ref.\cite{gam} but similar to that of Ref.\cite{lyn}.  The lower
critical field line,
separating the Abrikosov and Meissner phases, is estimated from the first
appearance of vortex scattering in the SANS.

For each ($T,H$), we measure the vortex SANS patterns for different thermal
paths.
For low temperatures the vortex SANS images show sharp Bragg peaks
 with sixfold symmetry
 in agreement with previous studies \cite{cri,lyn,gam},
 independent of the thermal history.
 An example is shown in the inset of Fig.1(b) for $H$=3.75 kOe and $T$=3.50 K.
However, the vortex SANS pattern starts to show striking history dependence
as the peak-effect regime is approached.  For clarity, we define the
field-cooled (FC) state as when
 the sample is cooled
 to the target temperature in a magnetic field,
 while the zero-field-cooled (ZFC) state is reached by cooling the sample
in zero field
 to the target temperature and then increasing the magnetic field to
the desired value.  A field-cooled-warming (FC-W) state is when the system
is cooled in field
to a much lower temperature ($\sim$ 2 K) then warmed back to the final
temperature.

For the FC path, the vortex SANS patterns
show nearly isotropic rings for $T_p<T<T_{c2}$ and broad Bragg spots for
$T<T_p$.
There is no clear sharpening in the Bragg peaks when $T_o$ is crossed.
Only at a lower $T<T_p$ do the Bragg peaks become sharp.  In contrast, for
the ZFC and
the FC-W paths, the sharp Bragg spots are observed for all temperatures up
to $T_{c2}$.
  Shown in
the top panel of Fig.2 are the ZFC and FC images at $H$=3.75 kOe
and $T$=4.40 K, which is just below $T_o$(3.75kOe)=4.50 K.
The images in the mid panel are
for $H$=4.00 kOe and $T$=4.40 K, which is 0.10 K above
$T_p$(4.0kOe)=4.30 K.
The intensities at the radial maximum
for the mid panel SANS data are
plotted in the lower panel. The sharp Bragg spots for the ZFC state
indicate a vortex lattice with long-range-order (LRO) \cite{gia},
 while the broad spots for the FC state suggest
 a disordered phase with short-range-order.

The orientational order of the vortex assembly
can be quantified by
the azimuthal widths of the Bragg peaks at the radial position of
the intensity maximum.
The azimuthal widths $\Delta \theta$ of the Bragg peaks
 can be obtained by fitting six
Gaussian peaks to the data, for each ($T,H$) and path.
Likewise, the translational order of the vortex assembly can be quantified
by the radial widths $\Delta Q$ of the Bragg peaks.
The Gaussian widths are plotted in Fig.3.
Clearly, the ZFC states are more ordered than the FC states,
translationally and orientationally, across the peak-effect regime.
To determine the positional
correlation of the vortex lines along the field direction, the rocking
curves (Bragg peak intensity vs. the relative angle between
the neutron beam and the applied
magnetic field) of the ZFC and FC states at $H$=3.75 kOe
and $T$=4.40 K are also measured. The rocking-curve width for the FC state
 is 30$\%$ larger
than that of the ZFC state, suggesting that the vortex lines
are entangled \cite{nel}
in the FC states, but nearly straight in the ZFC (and FC-W) states.

Since a simultaneous broadening in radial and azimuthal widths is
characteristic of a liquid (or glass) \cite{chai},
the hysteresis in both $\Delta \theta$ and $\Delta Q$
suggests a first-order vortex solid-liquid (or glass) transition.
A controversial issue is
the location of the underlying equilibrium phase transition in the vortex
matter
relative to the position of the peak effect \cite{bha,ala,lmv,tan}.
One interpretation \cite{bha,ala,lmv}
places the conjectured vortex solid-liquid transition $T_m$ at $T_p$,
consistent with the recent experiments \cite{ish,shi} in YBCO
where the magnetization jump was found to coincide with the peak effect
(the two are also related in other high-$T_c$ systems \cite{kha}).
Another widely held view, is based on the classical Lindemann criterion
which would place $T_m$ at $T_{c2}(H)$ for Nb, provided
the vortex-lattice elastic moduli are well-behaved \cite{hou}.
In this scenario,
the FC disordered phase seen here (as well as in \cite{lyn,gam}) is a
supercooled liquid and
the thermodynamic ground state is an ordered solid
across the entire peak-effect regime.  The third scenario
places $T_m$ at or below the onset of the peak effect \cite{tan}.

We find that it is possible to experimentally determine the ground state,
and consequently the approximate
position of $T_m$,
of the vortex system.   For this purpose, we use the
ac susceptibility coil to apply a small ac
magnetic field to shake the
vortex assembly and use the SANS to observe how the structure of the system
evolves.
Even though the true ground state of the vortex system
may not have been reached in the time scale of the shaking experiment due
to random pinning,
the evolution of the diffraction patterns leaves little doubt regarding
the nature of the ground state.

As shown in Fig.4(a), after applying an ac field
of 3.3 Oe at 100 Hz (1 kHz gives similar results) for $10^3$ sec,
the ordered ZFC vortex lattice becomes completely disordered.
 Preliminary time-dependent
data show that the Bragg peaks start to disappear
within the first $10^2$ sec of the shaking experiment.
In contrast, no measurable
difference can be found for the disordered FC vortex states before and
after the ac field is applied (for up to 10$^4$ sec).
By using the same approach, we find that the FC disordered states for
$T<T_p$ are metastable and the ordered ZFC state is the ground state,
opposite to that for $T>T_p$.
In the $T<T_p$ regime, the metastability is obviously stronger since a much
larger ac field
is needed to change the metastable state.  We find that an ac field of 50 Oe
(at $\approx$ 0.1 Hz, using the superconducting magnet at non-persistent
mode) can crystallize
the disordered FC states at $T<T_p$.
The shaking effects of an ac magnetic field
were also observed
in transport \cite{mwr} and ac magnetization \cite{ban} in the peak-effect
region in
2H-NbSe$_2$,
and were interpreted as a structural
re-organization in the vortex matter, consistent with our direct SANS
observations here.

The metastable nature of the ZFC
ordered state for $T>T_p$ can also be observed in a field ramping experiment.
For example, at a field ramp (increase) rate of less than 5 Oe/sec,
the final vortex state (at $T>T_p$) is always ordered.
In contrast, a ramp rate larger than 40 Oe/sec always results in a
disordered state.
However, the field ramping experiment alone cannot rule out
a trivial possibility that the sample was heated
to above $T_{c2}(H)$ during the field ramping by the induced screening
current.
If this happens, after the field ramping stops,
the sample cools back to
the set temperature {\it in} field and the final state is actually a FC
state.
In the shaking experiment, the ac susceptibility of
 the sample is monitored and serves as an {\it in situ}
thermometer to ensure that the sample temperature never fluctuates to
above $T_{c2}(H)$ during the entire SANS run (1 hour).

Thus we conclude that, for $T>T_p$,
the ordered ZFC vortex lattice is a superheated state and the
ground state of the vortex system is a disordered vortex liquid,
while for $T<T_p$, the ground state
is a vortex crystal (or Bragg glass \cite{gia}), and the disordered FC state
is a supercooled vortex liquid.
A thermodynamic phase transition must have taken place in the region of
 the peak effect, $T_m \approx T_p$. This is consistent with the observations
 in YBCO where the magnetization jump
was found to coincide with the peak effect \cite{ish,shi}.
In principle, one should also observe a discontinuous change of $Q_o$
when the vortex solid melts.
Unfortunately, the expected shift in $Q_o$ due to
the density change
at the vortex solid-liquid transition
is below our resolution limit, as we estimate here.
From thermodynamic considerations \cite{zel}, using the Clausius-Clapeyron
relations,
the vortex density change at $T_m$ is of the order
$\Delta B=-\frac{4{\pi}{{c_L}^2}C_{66}}{T_mdH_m/dT}$ where $c_L$ is the
Lindemann number
and $C_{66}\approx {\frac{B_{c2}^2}{4\pi}}{\frac{b(1-b)^2}{8\kappa^2}}$ is the
vortex-lattice shear modulus, $b=B/B_{c2}$.
For Nb at $H$=4.00 kOe, $T_m \approx T_p$=4.30 K ($T_{c2}$=4.50 K),
$dH_m/dT$= -0.846 kOe/K,
and $C_{66} $$\approx$ 0.6x10$^4$ erg/cm$^3$ assuming a reasonable
 $\kappa \approx$1 and Lindemann number $c_L$=0.1,
 the upper bound for the jump $\Delta B$ at
melting is about 0.2 G.  For a triangular lattice before melting
$Q_o(B=H)=2\pi\sqrt{\frac{2}{\sqrt{3}}}\sqrt{\frac{H}{\phi_o}}$,
the expected $\Delta Q_o$ $\approx$ 3x10$^{-5}Q_o$ which is far
too small to be resolved by the present SANS instruments.
In Fig.4(b), the radial peak position $Q_o$ remains unchanged after
shaking-induced melting,
consistent with the above estimate.

 Another interesting feature in Fig.3 is that there appears to be a
characteristic temperature
below which the vortex state is insensitive to its thermal history.
For example, at $H$=3.75 kOe, the FC, ZFC, and FC-W states at 3.50 K
are all ordered.
The crossover temperature, $t \approx 0.85$ in Fig.3, is in fact very close
to the
melting line identified previously \cite{lyn} (which was based on the
measurements of the FC states).
We interpret this temperature as the lower limit for supercooling.

 A closer examination of the diffuse scattering rings in Figs.1$\&$2
suggests that they are not completely isotropic,
indicating some short-range correlations.  We attribute this short-range order
to the coupling of the vortex liquid to the atomic crystal lattice,
since experimentally one finds that
in the Bragg solid phase, the specific orientation of the
vortex lattice is always fixed to a certain crystallographic orientation of
the sample.
(This is the
case for all low-$T_c$ and high-$T_c$ systems \cite{kei}.)  In the supercooled
state (top-right in Fig. 2), the highly correlated vortex liquid \cite{lyn},
reminiscent of the hexatic liquid phase in 2D melting \cite{hal},
is also likely related to this coupling. 

We point out that our results
imply the absence of superheating
in conventional transport experiments with a large drive current.  This solves
a well-known puzzle  \cite{hen,lbp} in which the history dependence of the
nonlinear
resistance always vanishes at $T_p(H)$.
Only with extremely low drive
currents may one observe the subtle effects of superheating in transport
\cite{cha}.
The superheating of vortex solid contrasts sharply with that in regular
bulk solids
where surface melting prevents superheating \cite{lun}.
Surface melting of vortex solid
is likely suppressed by the surface barrier\cite{cha} in type-II
superconductors.

In summary, we have observed a striking hysteresis in the SANS patterns of
the vortex matter
in the peak effect regime of a high quality Nb single crystal.
The FC vortex states are disordered across the entire peak effect region
and can be supercooled
to below the onset of the peak effect.  In contrast, the ZFC or the FC-W
vortex states
are ordered up to $H_{c2}(T)$.  The ordered ZFC/FC-W vortex lattices
near $H_{c2}(T)$ are identified as
superheated solids which can be melted (disordered) by applying a small ac
magnetic field.
We interpret our SANS hysteresis data as direct evidence for a first-order
vortex solid-liquid (glass)
transition.

We thank C.J. Glinka for assistance and J.J. Rush for providing the Nb
crystal;
C. Elbaum, A. Houghton, J.M. Kosterlitz, H.J. Maris, and S.C. Ying for
discussions.
This work was supported in part by NSF through Grant No. DMR-0075838, the
A.P. Sloan Foundation (XSL),
and the Brown Salomon Award program.

\newpage
{\bf Figure Captions:}

Fig.1: Peak effect and the field-temperature ($H-T$) phase diagram of Nb.
(a)
Representative ac magnetic susceptibility $\chi'(T)$ vs.
$T$ for $H_{dc}$=3.75 kOe (field-cooled).
$H_{ac}$=3.3 Oe and $f$=1.0 kHz.  $H_{dc}||H_{ac}$.
(For details on ac magnetic measurements, see \cite{lin,shi}.)
Inset: the global $H-T$ phase diagram for the Nb crystal used in this study.
(b) Expanded view of the $H-T$ phase diagram for Nb (the shaded box in (a)).
Two representative SANS images of the
field-cooled vortex states are shown (no background subtraction). \ \\

Fig.2: History-dependent neutron diffraction patterns. $T$=4.40 K for all
images.
 The SANS images of the ZFC and FC vortex states for $H$=3.75 kOe
 (top panel: below the onset of the peak effect) and $H$=4.00 kOe (mid
panel: near the upper end
 of the peak-effect regime).  The thick arrows indicate how the SANS images
evolve
 after applying a small ac magnetic field (see text).  The lower panel shows
the intensity data at the radial maximum as a function of the azimuthal
angle $\theta$ for
the ZFC and FC SANS data (background subtracted) at $H$=4.00 kOe and
$T$=4.40 K.
 \ \\

Fig.3: Temperature dependence of the SANS widths.
(a) The azimuthal widths of the Bragg peaks at $Q_o$ and (b)
the radial widths vs. reduced temperature for ZFC, FC, and FC-W vortex
states.
The FC-W data points at 3.75 kOe and 4.40 K are indicated by
the crosses (and marked by arrows).  The instrumental resolution in width
is marked in (a).
The lines are guides for the eye.  The dashed lines indicate $T_p(H)$.
\ \\

Fig.4: The effect of an ac field, $H_{ac}$=3.3 Oe and $f$=100.0 Hz,
on the ordered ZFC state for
$H$=4.00 kOe and $T$=4.40 K (above $T_p(H)$). (a) $I(Q_o)$ vs. azimuthal angle;
(b) the intensity $I(Q)$ (averaged over the azimuthal angle)
as a function of the momentum transfer $Q$.
The location of the radial maximum $Q_o$ is found by
fitting $I(Q)$ to a Gaussian. The lines are Gaussian fittings.

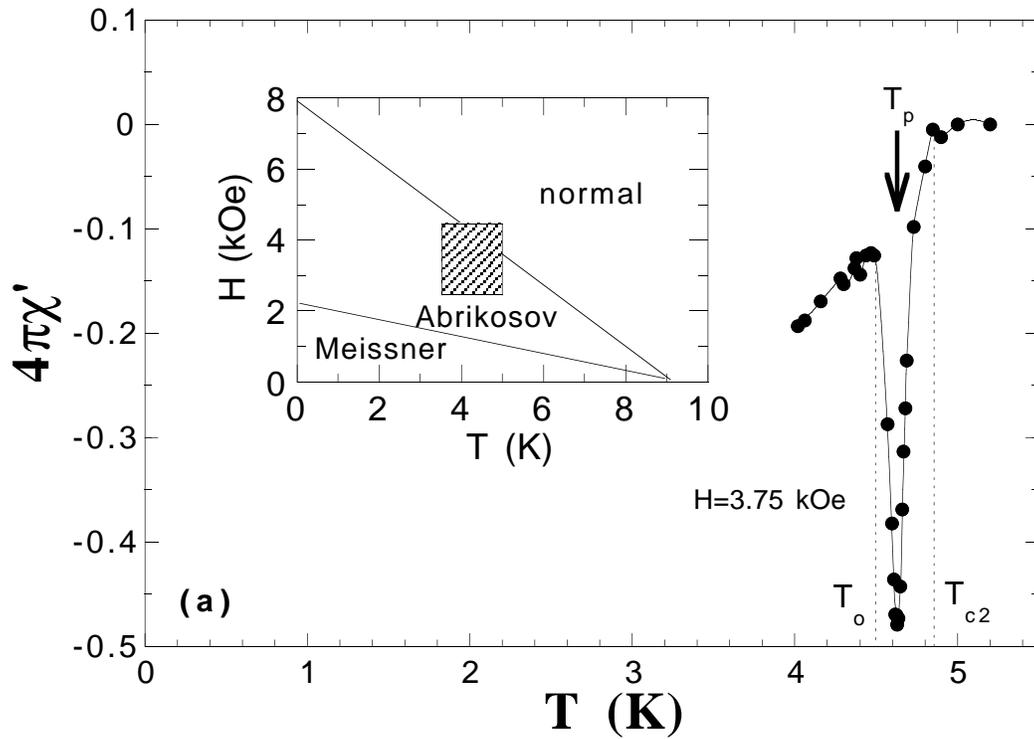
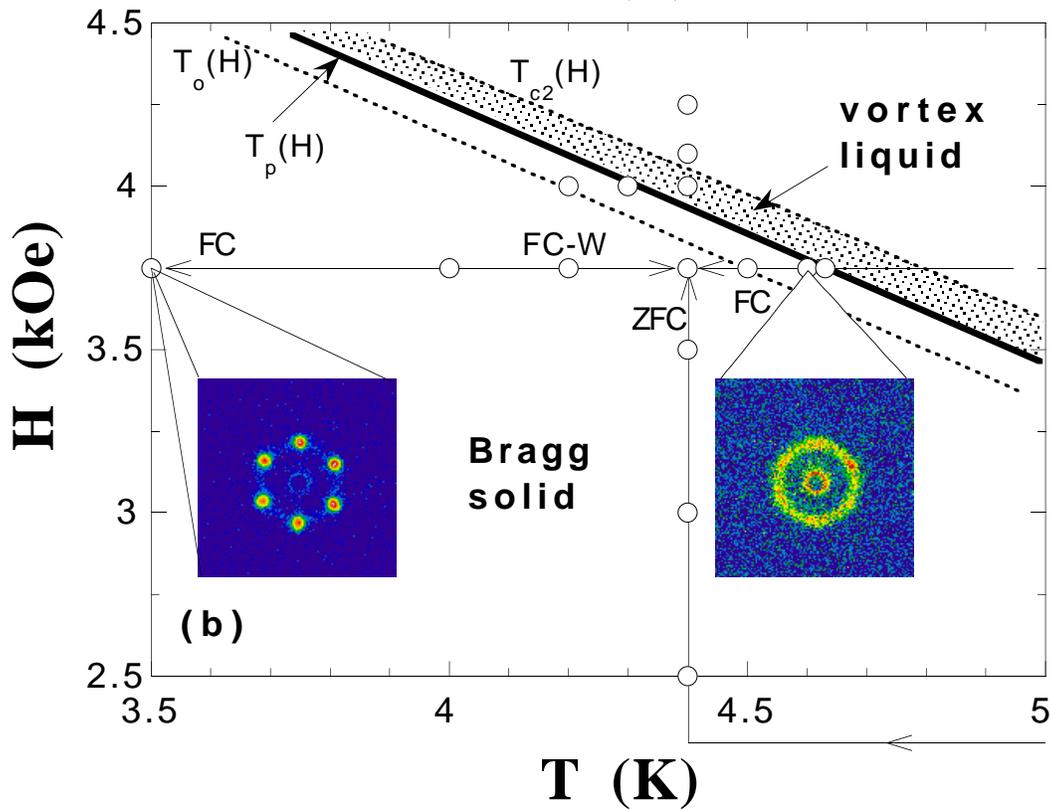
\end{document}